\documentclass[aps,prl,twocolumn,groupedaddress,superscriptaddress,nofootinbib]
{revtex4}
\usepackage{graphicx} 
\usepackage{dcolumn} 
\usepackage{longtable} 
\usepackage{amssymb} 
\usepackage{amsmath} 
\usepackage{amsfonts} 
\usepackage{slashed} 
\usepackage[dvipsnames]{xcolor} 
\usepackage{soul}

\newcommand{\be}{\begin{equation}} 
\newcommand{\ee}{\end{equation}}

\newcommand{\quotes}[1]{``#1''}

\begin{document}

\title{Limits on an improved action for contact effective field theory \\
in two-body systems}  

\author{Lorenzo Contessi} 
\affiliation{Université Paris-Saclay, CNRS-IN2P3, IJCLab, 91405 Orsay, France}
\author{Manuel Pavon Valderrama}
\affiliation{School of Physics, Beihang University, Beijing 100191, China} 
\author{Ubirajara van Kolck}
\affiliation{European Centre for Theoretical Studies in Nuclear Physics and Related Areas (ECT*), Fondazione Bruno Kessler, 
38123 Villazzano, Italy}
\affiliation{Université Paris-Saclay, CNRS-IN2P3, IJCLab, 91405 Orsay, France}
\affiliation{Department of Physics, University of Arizona,
Tucson, AZ 85721, USA}
\date{\today}

\begin{abstract} 
  We consider a possible resummation of subleading effects in two-body systems with a large scattering length as described by a   short-range effective field theory (EFT). In particular, we investigate the consequences of a resummation of part of the range corrections. Explicit calculations of the two-body phase shifts and charge form factor indicate that, except for extreme choices, resummations do not alter the convergence of the EFT expansion and are often beneficial at lowest orders. We have considered the expansion when the regulator cutoff is removed as well as when it is finite, and find that the cutoff is not an important factor for resummations. Our results connect with other works where the partial resummation is induced by potentials with finite cutoffs or interaction ranges.
\end{abstract}

\maketitle

\noindent 
{\bf Introduction: }
Effective field theories (EFTs) are general low-energy descriptions of physical systems. They are formulated as a power series on a small parameter, a property known as power counting. Independence on the details of the underlying dynamics demands renormalization order by order. Conventional wisdom states that a well-defined power counting requires that subleading orders in the EFT series be treated as perturbative corrections, even in systems where leading order (LO) is nonperturbative so as to produce bound states and resonances~\cite{Hammer:2019poc}. Yet, this might be cumbersome, particularly in few-body calculations. Thus, it would be useful to know under which circumstances the nonperturbative treatment or resummation of subleading-order contributions does not upset the renormalizability of the EFT and its power counting.

Because more straightforward, the nonperturbative treatment of subleading corrections has become standard in nuclear physics with potentials and currents~\cite{Epelbaum:2008ga,Machleidt:2011zz} derived from Chiral EFT~\cite{Hammer:2019poc}. Though naively it could be argued that subleading interactions --- whose effects on observables are by definition small --- should still be subleading after iteration, reality does not necessarily comply with this expectation. The most evident problem is that the resummation of subleading orders may alter how the EFT is renormalized, as has been explicitly shown for chiral two-pion exchange~\cite{PavonValderrama:2005wv,PavonValderrama:2005uj}. In this concrete case, resummations are incompatible with the formulation of a power counting, which requires the explicit perturbative treatment of the subleading order corrections~\cite{Long:2007vp,Valderrama:2009ei,Valderrama:2011mv,Long:2011qx,Long:2011xw,Long:2012ve}, as conjectured in Ref.~\cite{Nogga:2005hy} and reviewed in Ref.~\cite{vanKolck:2020llt}.

This breakdown of an expectation based on experience with regular potentials is a consequence of the fact that EFT interactions can be singular already at LO, and become more singular at higher orders. The incompatibility of indiscriminate resummation of subleading interactions with renormalization was first pointed out in Refs.~\cite{Phillips:1996ae,Beane:1997pk} in the context of Pionless (or more generally Short-Range) EFT~\cite{Hammer:2019poc}. In this case, the resummation of momentum-dependent interactions in the two-body system leads to the Wigner bound~\cite{Wigner:1955zz}, which restricts the effective range that can be reproduced with contact interactions to be negative, $r_0\le 0$. While this situation is relevant for low-energy resonances~\cite{Habashi:2020qgw,Habashi:2020ofb,vanKolck:2022lqz}, it conflicts with many systems of interest, such as the two-nucleon system, where $r_0>0$.

As long as the effective range is comparable to the range of the underlying interaction, it can be treated perturbatively. However, at LO both Pionless~\cite{Stetcu:2006ey,Contessi:2017rww,Bansal:2017pwn} and Chiral~\cite{Yang:2020pgi} EFTs tend to produce unstable nuclei beyond the alpha particle. Even though poles might exist close to threshold~\cite{Contessi:2022vhn}, it is not clear if and how these poles become stable at subleading orders. One possibility is the necessity of a genuine change in power counting~\cite{Yang:2021vxa}. Another is a resummation of specific subleading corrections into an improved LO action, which does not put renormalization and power counting at odds, if it only involves specific contact-range or non-singular finite-range interactions.

In this manuscript, we study a {\it partial} resummation of range effects in weakly bound two-body systems, which can be addressed with Short-Range EFT~\cite{Hammer:2019poc}. This approach has recently been shown to be beneficial in few-body systems~\cite{Contessi:2023yoz} when implemented through a local auxiliary potential in coordinate space, with an artificial range at LO which is removed systematically at subleading orders. We investigate here the extent to which effective-range effects can be resummed using simpler, energy-dependent interactions. In this case, there are no renormalization obstacles, as the Wigner bound, to the resummation of terms in the effective-range expansion, allowing us to focus on the convergence of the theory with momentum, both before and after the removal of a regularization cutoff. By comparing calculated scattering phase shifts and a bound-state form factor with results from a toy underlying theory, we will show that power counting is not fundamentally affected, as both resummed and nonresummed expansions have similar convergence properties. We can thus capture some of the subleading effects at the lowest orders. A resummation of the full effective range has already shown improvement at two-body level \cite{Phillips:1999hh}, but improvement is not restricted to this case, as expected from the fact that the artificial range is not a physical input at LO.

\vspace{5mm}
\noindent
{\bf Toy underlying theory: }
In the absence of shallow poles or resonances in higher waves, the low-energy scattering between two particles of mass $m$ and center-of-mass momentum $k\ll m$ is dominated by the $S$ wave. The phase shift $\delta(k)$ is given by the effective range expansion (ERE)
\begin{eqnarray}
    k\cot \delta(k)= -\frac{1}{a_0} +\frac{r_0}{2} \, k^2 + v_2 \, k^4 + v_3 \, k^6 +\ldots
\end{eqnarray}
in terms of the scattering length $a_0$, effective range $r_0$, and shape parameters $v_{n}$, $n=2,3,\ldots$. Introducing 
\begin{equation}
    K\equiv \frac{4\pi a_0}{m} \, ,
\end{equation}
and the dimensionless parameters
\begin{equation}
    \eta\equiv k a_0, 
    \quad
    \alpha\equiv \frac{r_0}{2a_0},
    \quad
    \nu_n\equiv \frac{v_n}{a_0^{2n-1}},
\end{equation}
the on-shell $T$ matrix can be written as
\begin{equation}
  T(\eta) = - K \left[\eta\,\cot\delta(\eta) - i\,\eta \right]^{-1} \, , 
  \label{eq:T-matrix_delta}
\end{equation}
with 
\begin{equation}
    \eta \cot \delta(\eta)= -1 +\alpha \, \eta^2 + \nu_2 \, \eta^4 + \nu_3 \, \eta^6 +\ldots
    \label{eq:dimlesscotdelta}
\end{equation}

The ERE holds generally for $k\ll R^{-1}$, where $R$ is the range of potential. We are interested in systems, like those of two nucleons or two $^4$He atoms, where the $T$ matrix has a single low-energy pole, either a bound or a virtual state at an imaginary momentum $k=i\kappa$, $|\kappa|\ll R^{-1}$. In this situation, the potential is fine-tuned to yield $\alpha\ll 1$, $\nu_n\ll 1$. Alternatively to Eq. (\ref{eq:T-matrix_delta}), the amplitude can be expanded around the pole momentum,
\begin{equation}
  T(\eta') = - K' \left[\eta'\,\cot\delta(\eta') - i\,\eta' \right]^{-1} \, , 
  \label{eq:T-matrix_delta_pole}
\end{equation}
with
\begin{equation}
    K'\equiv \frac{4\pi}{m\kappa} \, ,
\end{equation}
and
\begin{eqnarray}
  \eta'\,\cot{\delta(\eta')}&=&
  - 1 + \alpha' \left(\eta'^2 + 1\right)
  + V'_2\left(\eta'^4 - 1\right)
  \nonumber \\
  && + V'_3\left(\eta'^6 + 1\right)
  + \dots \, ,
  \label{eq:ERE-pole}
\end{eqnarray}
where
\begin{equation}
    \gamma\equiv \kappa \,a_0,
    \quad 
    \eta'\equiv \frac{\eta}{\gamma},
    \quad
    \alpha'\equiv \alpha \,\gamma,
    \quad
    \nu'_n\equiv \nu_n \gamma^{2n-1}.
\end{equation}

Specific finite-range potentials might produce a truncated version of the ERE at all $k\ll m$. Examples of potentials that give a truncation at ${\cal O}(\eta^2)$ are given, for example, in Ref. \cite{RevModPhys.21.488}. In what follows we will consider the less particular case of a toy underlying theory where the potential is such that the ERE is exact when truncated at ${\cal O}(\eta^6)$. In other words, the phase shifts $\delta(\eta)\equiv \delta_{\rm toy}(\eta)$ are obtained from Eq. \eqref{eq:dimlesscotdelta} with
\begin{eqnarray}
\nu_4 = \nu_5 = \ldots = 0 \, .
\label{eq:Vn>3=0}
\end{eqnarray}

This underlying theory is defined by the three parameters $\alpha$, $\nu_2$, and $\nu_3$.  For concreteness, we set the ERE parameters to those of one of the relevant cases, namely, spin-triplet neutron-proton scattering, which we take from Ref.~\cite{deSwart:1995ui}. (Essentially the same results are obtained~\cite{PavonValderrama:2004se} from the Nijmegen II and Reid 93 potentials~\cite{Stoks:1994wp}.) The scattering length is $a_0=5.420$ fm and the dimensionless ERE parameters are summarized in Table \ref{tab:dimlessparam}. With these parameters, the truncated ERE admits a bound state --- the deuteron of our underlying theory --- at positive imaginary momentum, 
$\kappa>0$, whose dimensionless value is also given in Table \ref{tab:dimlessparam}. 

\begin{table}[t]
\def\arraystretch{1.5}
\setlength\tabcolsep{6pt}
\begin{center}
\begin{tabular}{ccc|c}
\hline
$10 \, \alpha$ & $10^{4}\, \nu_2$ & $10^{4}\, \nu_3$ & $\gamma$\\
\hline
1.617 & 2.512 & 1.437 & 1.254
\end{tabular}
\end{center}
\caption{Values used in this work for the dimensionless half-effective-range range $\alpha=r_0/2a_0$ and shape parameters $\nu_{2}=v_2/a_0^3$,
$\nu_{3}=v_3/a_0^5$, as well as the resulting binding momentum $\gamma=\kappa\,a_0$.}
\label{tab:dimlessparam}
\end{table}

The corresponding scattering matrix $T_{\rm toy}(\eta)$ represents our toy theory, which we will reproduce within the improved-action EFT approach. Our results can easily be extended to other values of ERE parameters and a truncation of the ERE at higher powers of $\eta^2$. Since there is nothing special about a truncation at ${\cal O}(\eta^6)$, our qualitative conclusions should hold for a generic finite-range potential that yields a single low-energy $S$-wave pole. 

\vspace{5mm}
\noindent
{\bf EFT:}
Short-Range EFT is designed to capture the dynamics at $k\ll R^{-1}$ in a systematic expansion. The usual power counting for a two-body system with an unnaturally large scattering length~\cite{vanKolck:1998bw,Chen:1999tn} is defined for $k\sim a_0^{-1}$ by counting powers of $QR$, where $Q = \{a_0^{-1}, k \}$ represents the low-energy scales. On the basis of dimensional analysis, we expect for a generic potential that $\alpha={\cal O}(QR)$ and $\nu_n={\cal O}(\alpha^{2n-1})$. While the value of $\nu_{2}$ in Table \ref{tab:dimlessparam} is actually an order of magnitude smaller, that of $\nu_3$ conforms to expectation. At LO, which corresponds to a point interaction ($R\to 0$), $K$ is the only parameter. The range of the interaction enters at subleading orders in a distorted-wave expansion of the amplitude in powers of $QR$, with $\alpha$ and $\nu_n$ first entering at, respectively, NLO and N$^{2n-1}$LO. 

Strict implementation of this power counting yields good results for few-body systems near the two-body unitarity limit~\cite{Hammer:2019poc}, but tends to produce unstable systems of multiple multi-component fermions in the point-interaction limit. This suggests that we improve the action to allow part of the effective range $r_0$ to be iterated together with the LO interaction responsible for the energy-independent part of the $T$ matrix. For this purpose, we assume energy-dependent interactions that avoid the Wigner bound~\cite{Wigner:1955zz}. In terms of a real number $x$, we write
\begin{equation}
    \alpha= x \alpha + (1-x) \alpha
    \label{eq:x}
\end{equation}
and treat $x\alpha$ nonperturbatively. This would be necessary for $x\sim 1$ when $|\alpha|\sim 1$, but with energy-dependent interactions it can be done for any $x$ without renormalization problems when $|\alpha|\ll 1$. 

We now check whether this modification leads to inconsistencies in the EFT expansion or changes in the convergence radius. The EFT energy-dependent contact interactions are fitted to yield the expansion of the toy $T$ matrix,
\begin{eqnarray}
  T(\eta)
  = \sum_{n=0}^{\infty} T^{(n)}(\eta) \, , 
  \label{eq:T-exp}
\end{eqnarray}
with
\begin{align}
  T^{(0)}(\eta)=& - K\, \left(-1+x\,\alpha\,\eta^2-i\,\eta\right)^{-1}\, ,
  \label{eq:power_LO} \\
  T^{(1)}(\eta)=& - K^{-1}\,\left[T^{(0)}(\eta)\right]^2\, (1-x)\,\alpha\,\eta^2 \, ,
  \label{eq:power_NLO} \\
  T^{(2)}(\eta)=& \, K \left[T^{(1)}(\eta)\right]^2\left[T^{(0)}(\eta)\right]^{-1}
  \label{eq:power_N2LO} \\
  T^{(3)}(\eta)=& - K^{-1}\,\left[T^{(0)}(\eta)\right]^2\, \nu_2\,\eta^4  + \dots \, , 
  \label{eq:power_N3LO} \\
  T^{(4)}(\eta)=& \, 2K \, T^{(3)}(\eta) \, T^{(1)}(\eta)\left[T^{(0)}(\eta)\right]^{-1}
  + \dots\, ,
  \label{eq:power_N4LO} \\
  T^{(5)}(\eta)=& - K^{-1}\,\left[T^{(0)}(\eta)\right]^2\, \nu_3\,\eta^6  + \dots \, .
  \label{eq:power_N5LO}
\end{align}
Here, at each order we show only the terms with smallest powers of $T^{(0)}$. These expressions can be obtained by introducing an auxiliary dimer field \cite{Kaplan:1996nv}, where the $x\alpha$ contribution in $T^{(0)}$ comes from the dimer kinetic term. The regulator dependence at LO can be removed by the dimer mass, and by other dimer parameters at subleading orders. With a regulator with momentum cutoff $\Lambda$, there remains a residual dependence in powers of $\Lambda^{-1}$, which vanishes as $\Lambda\to \infty$. The expressions for $T^{(n)}$ reduce for $x=0$ to the standard Short-Range EFT $T$ matrix~\cite{Hammer:2019poc} under the condition \eqref{eq:Vn>3=0}. 

The phase shift of the toy theory, $\delta_{\rm toy}$, is calculated by considering its relation to the $T$ matrix via Eq. \eqref{eq:T-matrix_delta}. In contrast, the EFT phase shift is expanded in powers of $QR$,
\begin{eqnarray}
  \delta
  (\eta)
  = \sum_{n=0}^{\infty} \delta^{(n)}(\eta) \, , \label{eq:delta-exp}
\end{eqnarray}
and obtained from the corresponding expansion for the $T$ matrix, Eq.~(\ref{eq:T-exp}):
\begin{eqnarray}
  \cot{\delta^{(0)}(\eta)} &=&
  -\frac{K}{\eta}\,{\rm Re}\left(T^{(0)}(\eta)\right)^{-1} \, , 
  \label{eq:EFTPS0}\\
  \frac{\delta^{(1)}(\eta)}{\sin^2{\delta^{(0)}(\eta)}} &=& - \frac{K}{\eta}\,
  \left[{\rm Re}\left(T^{(0)}(\eta)\right)^{-1} \right]^2\,T^{(1)}(\eta) \, ,  \nonumber \\
  \label{eq:EFTPS1}
\end{eqnarray}
plus the expressions for higher orders, which are lengthy but straightforward.

\begin{figure*}
  \centering
  \begin{tabular}{cc}
    \multicolumn{2}{c}{
      \includegraphics[width=0.6\textwidth]{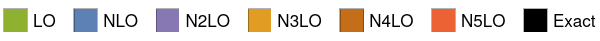}} \\
    \includegraphics[width=0.45 \linewidth]{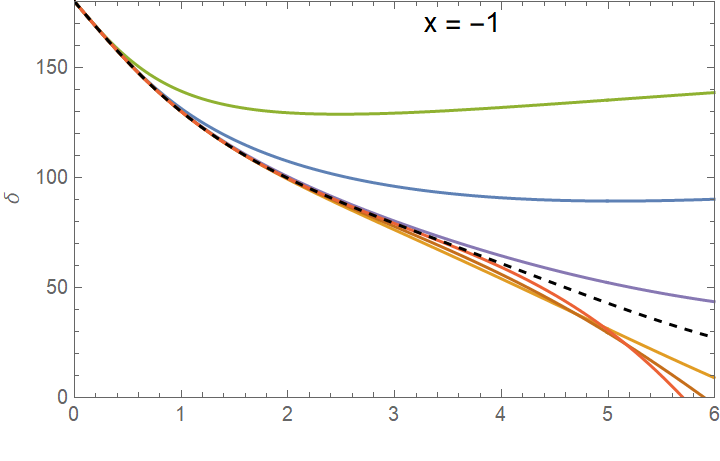} &
    \includegraphics[width=0.45 \linewidth]{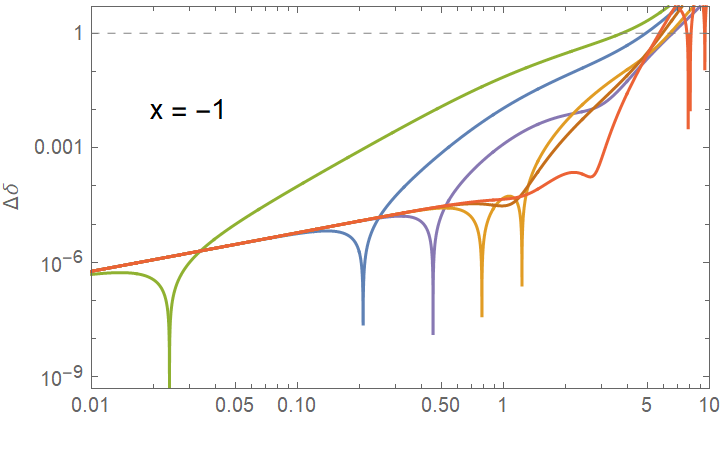}   \\
    \includegraphics[width=0.45 \linewidth]{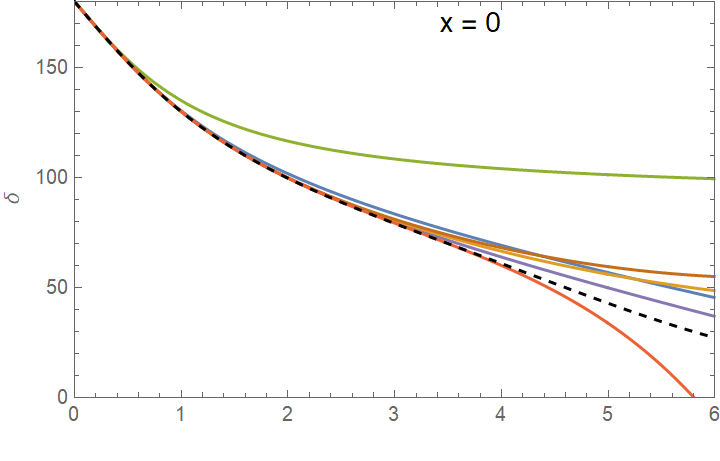}  &
    \includegraphics[width=0.45 \linewidth]{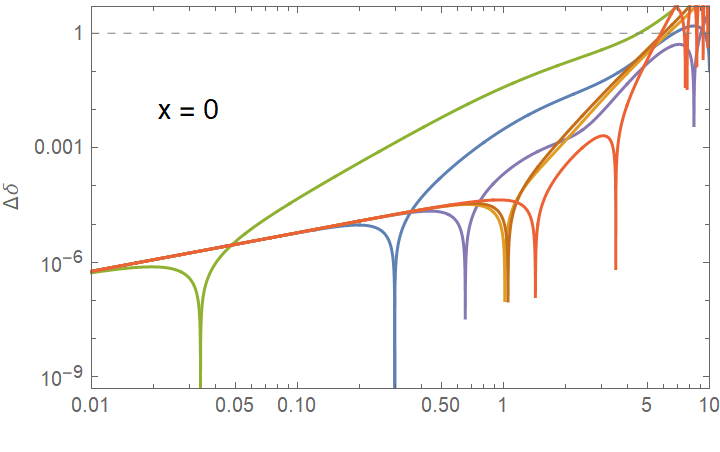}   \\
    \includegraphics[width=0.45 \linewidth]{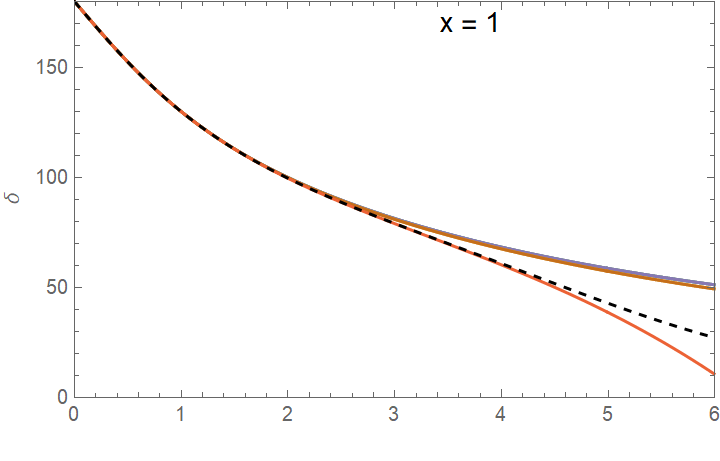}  &
    \includegraphics[width=0.45 \linewidth]{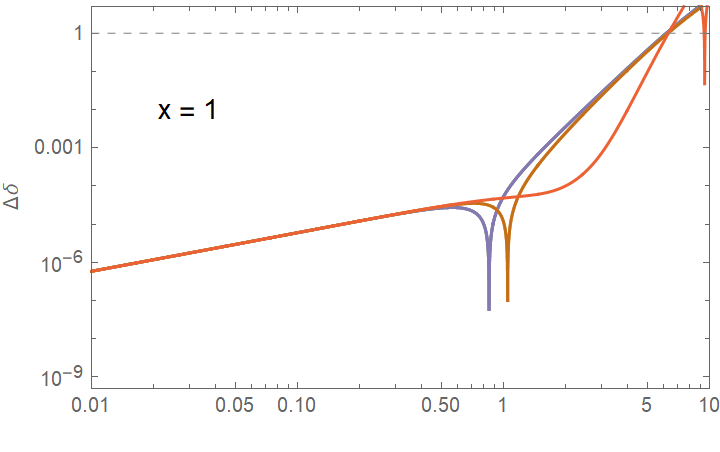}   \\
    \includegraphics[width=0.45 \linewidth]{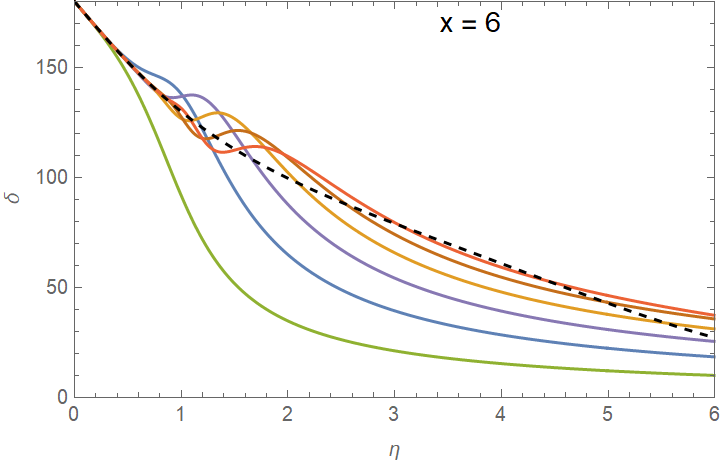} &
    \includegraphics[width=0.45 \linewidth]{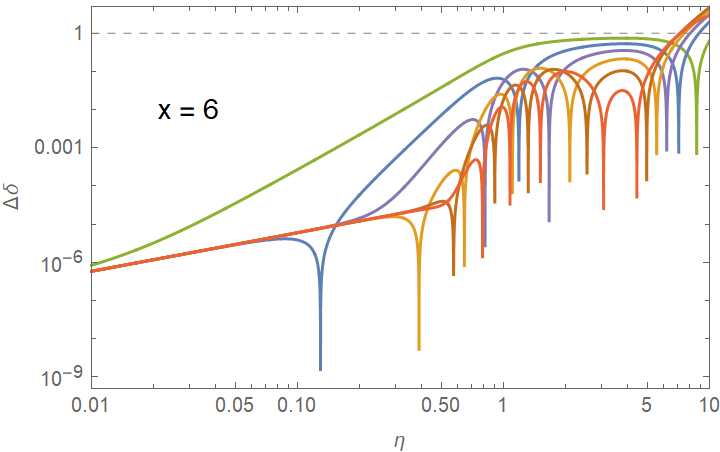} \\  
  \end{tabular}
  \caption{Improved EFT expansion of the phase shifts up to N$^5$LO   for selected values of the resummation parameter $x$: $x=-1$ (top row), $x=0$ (second row), $x=1$ (third row), and $x=6$ (bottom row). On the left panels, EFT and toy-theory phase shifts (in degrees) are shown as functions of the reduced momentum $\eta$. On the right panels, the differences between EFT and toy phase shifts divided by the toy phase shift ($\Delta \delta$) are shown as functions of $\eta$, both on logarithmic scales. LO, NLO, N$^2$LO, N$^3$LO, N$^4$LO, and N$^5$LO results are indicated by green, blue, purple, light orange, brown, and orange solid lines, respectively, while toy-theory results are given on the left panels by a black dashed line.
  }
  \label{fig:table_of_phaseshifts}
\end{figure*}

\vspace{5mm}
\noindent
{\bf Comparison of amplitudes: }
In Fig.~\ref{fig:table_of_phaseshifts} we present the results of the EFT expansion up to N$^5$LO, for the values of ERE parameters given in Table \ref{tab:dimlessparam} and for a range of values of the dimensionless resummation parameter $x$. In the left panels, we display the EFT phase shifts in comparison with the phase shifts from the toy underlying theory. Visual inspection suggests that in all cases there is convergence to the toy-theory results as order increases. For a finer analysis, in the right panels, we show the corresponding Lepage plots~\cite{Lepage:1997cs}, log-log plots for the difference between the toy phase shift and its truncated value, as defined by Eq.~(\ref{eq:delta-exp}), divided by the toy value. These plots show how the error of the EFT expansion scales with momentum. At each order, the error increases with a power corresponding to the order of the neglected contributions. As the next order is considered, the slope increases because higher powers of momentum are accounted for. The overall error decreases, up to the EFT breakdown scale. The cusps represent momenta in which the EFT phase shifts happen to cross the toy data. The region where the lines cross provides an estimate of the convergence radius of the theory: the momentum in which there is no refinement in observables as order increases. 

For the unimproved EFT expansion, $x=0$, we see the expected systematic pattern of error decrease with clearly separated curves, except when going from N$^3$LO to N$^4$LO. This is due to the abnormally small value of the shape parameter $\nu_2$ in Table \ref{tab:dimlessparam}. The expansion converges at least up to a reduced momentum of $\eta \approx 4.5$, beyond which the ${\rm N^5LO}$ result starts to deviate from the toy-theory  result. Since the model is already truncated at ${\cal O}(\eta^6)$ of the ERE, any difference with the expansion up to ${\rm N^5LO}$ arises from EFT truncation errors rather than from missing ERE parameters.

For $x\ne 0$, the pattern of error improvement at lower orders can change considerably. The extreme case is that of including the full, physical effective range $r_0$ at LO, $x=1$, when its contributions at NLO, N$^2$LO, and N$^4$LO vanish. The convergence rate of the improved EFT expansion is greatly enhanced up to $\eta\approx3$, after which the $\nu_2$ and $\nu_3$ contributions become essential and the overall convergence is similar to 
that of the usual, unimproved expansion.

As $x$ increases further and we include in the improved LO range effects larger than physical, lower-order results deteriorate. A continuous transition takes place to what is observed for $x=6$, where oscillations appear at low energies ($\eta\approx1.2$) and deviations from the toy-theory value are seen at large momenta. Deterioration is also found for $x<0$, although when $x=-1$ the convergence pattern is still similar to $x=0$. Negative $x$ is interesting because the improved theory can be renormalized with momentum-dependent interactions \cite{Phillips:1996ae,Beane:1997pk}. However, $x<0$ is unlikely to help with the LO stability problem in many-body systems.

Remarkably, while the convergence rate of the lower orders does indeed depend on the resummation parameter $x$, the higher orders of the expansion are little affected. The convergence radius of the theory is mostly unaffected by the choice of $x$, a fact that indicates that there is no harm in the resummation of most of the range corrections.

This is easy to understand when we notice that the higher-order 
ERE parameters $v_2$ and $v_3$, which are not resummed, also contain information about the hard scale of the EFT. Thus, while the resummation of effects coming from $r_0$ can indeed accelerate the expansion at lower orders, the shape parameters begin to appear at high enough orders. This eventually restores the previous convergence rate of the EFT expansion prior to the resummation of $r_0$ effects. Of course, further improvement for this particular toy model could be achieved by additional resummation of $\nu_2$ and $\nu_3$. However, in a more general situation in which the ERE expansion is not truncated the same problem would reoccur at higher EFT order. Although practical, resummation does not seem to alter the convergence radius of the EFT.

The improvement is useful as long as $x$ is not very large, when it worsens also the convergence radius. Since in our case $\alpha^{-1}\simeq 6$, for $|x|\sim 6$ we are adding to LO a term of ${\cal O}(1)$, thus creating a second low-energy pole. In this case, Eq. \eqref{eq:power_NLO} is no longer a perturbative correction. We can think of the improved EFT expansion as an expansion in $|x\alpha|$ superimposed on the usual EFT expansion. The most significant effect enters already at NLO, whose relative magnitude is captured by the ratio
\begin{equation}
  \xi_{T}(x,\eta)= \left| \frac{T^{(1)}(\eta)}{T^{(0)}(\eta)}\right|=
  \left|\frac{(1-x)\alpha\eta}
       {1-x\alpha\eta^2+i\eta}\right| 
    \label{eq:convergence_T}
\end{equation}
for the $T$ matrix, or
\begin{align}
   \xi_{\delta}(x,\eta) = & \Bigg| \frac{\delta^{(1)}(\eta)}{\delta^{(0)}(\eta)}\Bigg|
  = \Bigg|\frac{(1-x) \alpha \eta^3}{
  \left(1-x\alpha\eta^2\right)^2+\eta^2}
  \nonumber\\ 
    &\times \left\{{\rm arccot}\left(\frac{1}{\eta}\left(-1+\alpha x \eta^2\right)\right)\right\}^{-1}
    \Bigg| 
    \label{eq:convergence_fac}
\end{align}
for the phase shift. 

\begin{figure}
  \centering
  \includegraphics[width=0.85\linewidth]{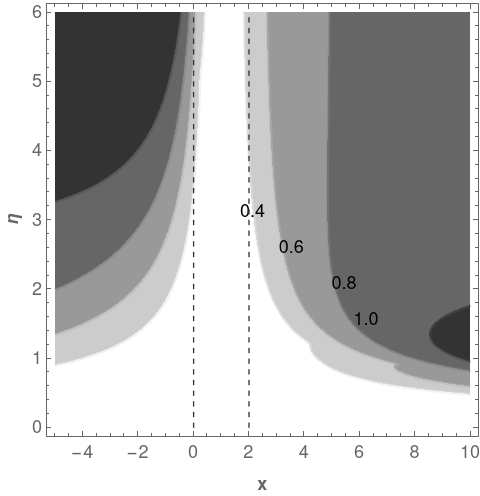} 
  \caption{Lowest-order convergence rate for the improved phase shifts, $\xi_{\delta}$ \eqref{eq:convergence_fac}, as a function of the reduced momentum $\eta$ and the resummation parameter $x$. The numbers at the borders between gray areas indicate the value of $\xi_{\delta}$ at the contour lines.
  }
  \label{fig:convergence}
\end{figure}

The behavior of the ratio $\xi_{\delta}(x,\eta)$ is shown in Fig.~\ref{fig:convergence}, where darker shades of gray represent larger relative NLO contributions. As expected, NLO is relatively small for a wide range of reduced momenta $\eta$ in the region $0 \lesssim x \lesssim2$. Outside this region, the ratio becomes larger for higher reduced momenta. Around $x \approx 8$ and $\eta\approx 1$ we have that $\xi_{\delta}\ge 1$, indicating a breakdown of perturbation theory. This is indeed consistent with the slow oscillatory convergence pattern we already saw at $x=6$ for the phase shifts in Fig.~\ref{fig:table_of_phaseshifts}. For $x<0$, the perturbative region is reduced. The interesting outcome of this analysis is the relatively large range of $x$ and $\eta$ where NLO remains perturbative. Similar conclusions hold for $\xi_T(x,\eta)$ in Eq.~\eqref{eq:convergence_T}.

Though not shown here, we have verified that qualitatively the same results are found with the ERE parameters from spin-singlet neutron-proton scattering~\cite{deSwart:1995ui,PavonValderrama:2005ku}: $10^2\, \alpha = -5.627$, $10^5\,\nu_2=3.563$, $10^7\,\nu_3=-5.269$, and 
$\gamma= 0.9493$,
with $a_0= -23.727$ fm. The main difference is that the deviation between toy and ${\rm N^5LO}$ results appears at larger momenta, namely $\eta \approx 6$. This suggests that the convergence radius of the expansion is relatively larger for the singlet channel.

\vspace{5mm}
\noindent
{\bf Finite-cutoff effects:}
The previous analysis exploited the fact that expressions for the two-body $T$ matrix of a short-range theory are known analytically when the regulator cutoff is removed. Yet, we can also consider the effects that a finite cutoff will have on the convergence of the EFT expansion, as in most situations calculations have to be performed at finite, albeit hard enough, cutoffs.

As an example, we consider the  case in which the two-body interaction is regularized with a delta-shell regulator at the radius $r = R_c$. For $a_0>0$, defining $\rho_c\equiv R_c/a_0=2\alpha R_c/r_0$, the phase shift is obtained from~\cite{Valderrama:2016koj}
\begin{eqnarray}
  && \eta\,\cot{(\delta(\eta) + \eta \rho_c)} - \eta\,\cot{(\eta \rho_c)} \nonumber \\ 
  && \qquad =
  c_0(\rho_c) + 
  c_1(\rho_c) \,\eta^2 
  + c_2(\rho_c)\, \eta^4 
  + \ldots 
  \, , \label{eq:kcot-finite-cutoff}
\end{eqnarray}
where $c_0$, $c_1$, ..., $c_n$ are reduced couplings, 
\begin{eqnarray}
  c_0(\rho_c) &=& \frac{1}{\rho_c(\rho_c - 1)} \, , \\
  c_1(\rho_c)&=& 
  \frac{1}{(\rho_c - 1)^2} \left[\alpha+ \frac{\rho_c\,(\rho_c-2)}{3}\right]
  \nonumber \\
  &\equiv& \frac{\alpha}{(\rho_c - 1)^2} + \Delta c_{1}(\rho_c)
  \, , \\
  c_{n\geq 2}(\rho_c)&=& \frac{\nu_{n}}{(\rho_c - 1)^2} + \Delta c_{n}(\rho_c)
   \, .
\end{eqnarray}
That is, $c_1$, $c_2$, ..., are chosen to reproduce $\alpha$, $\nu_2$, ..., respectively. The reduced couplings contain two contributions: the first is related to the ERE parameter they reproduce; and the second, which we show explicitly only for $c_1$, removes the cutoff dependence up to the order we are expanding, without adding new physical information. 

\begin{figure}
  \centering
  \includegraphics[height=5cm]{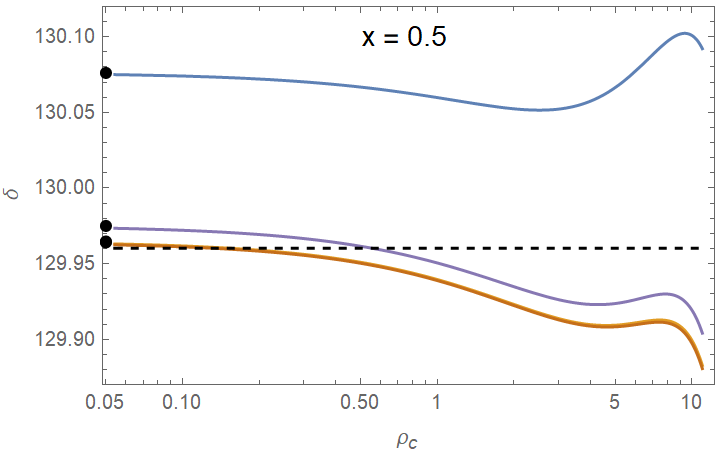} 
  \caption{Dependence of the phase shifts up to N$^5$LO in the improved expansion as a function of the dimensionless delta-shell regulator radius $\rho_c$ for $\eta=1$ and $x=0.5$. Curves as in Fig. \ref{fig:table_of_phaseshifts}. The LO result has been left out to enhance the visibility of the cutoff variation at subleading orders since the difference between LO and NLO is much greater than the cutoff variation. The solid circles on the left represent the values obtained previously from Eqs. \eqref{eq:EFTPS0}, \eqref{eq:EFTPS1}, {\it etc.}}
  \label{fig:cutoff}
\end{figure}

Our toy phase shift continues to be the one generated by Eq. \eqref{eq:Vn>3=0}, which means $c_{n\ge 4}(\rho_c)=\Delta c_n(\rho_c)$. In the EFT at N$^m$LO, we truncate Eq. \eqref{eq:kcot-finite-cutoff} with $c_{n>m}(\rho_c)=0$ and solve for the phase shift $\delta^{(m)}(\eta;\rho_c)$ in perturbation theory. In the improved expansion, we split $c_1(\rho_c)$ into two terms according to Eq. \eqref{eq:x}, with $x\alpha$ and $\Delta c_1$ treated nonperturbatively. The residual cutoff dependence of the phase shifts in both cases should disappear as the radius is decreased, $\delta^{(m)}(\eta;\rho_c\ll 2\alpha)\simeq \delta^{(m)}(\eta)$. An example is given in Fig. \ref{fig:cutoff} for momentum $\eta=1$ and a partial resummation of the effective range with $x=0.5$. We see that as the coordinate-space cutoff decreases the results converge to the phase shifts obtained in the contact limit from Eqs. \eqref{eq:EFTPS0}, \eqref{eq:EFTPS1}, {\it etc.}

A qualitatively similar picture was found for any other $\eta$ and $x$ studied. That is, the improved theory is always cutoff-convergent in the contact limit for all the cases considered, demonstrating that the resummation procedure does not alter the theory's renormalizability.

\vspace{5mm}
\noindent
{\bf Bound state:}
A large, positive scattering length signals the existence of a shallow bound state. In the vicinity of the pole, the EFT $T$-matrix expansion needs to be reorganized. At LO without improvement, the expansion contains multiple-pole terms at a binding momentum $\gamma^{(0)}=1$. In a region $|\gamma-\gamma^{(0)}|={\cal O}(QR)$ around $\gamma^{(0)}$, the formally NLO term must be resummed, displacing the pole to $\gamma^{(0)}+\gamma^{(1)}=1+\alpha$. The procedure continues at higher orders in ever shrinking regions, with successively smaller displacements. 

\begin{figure}[tb]
  \centering
  \includegraphics[height=5cm]{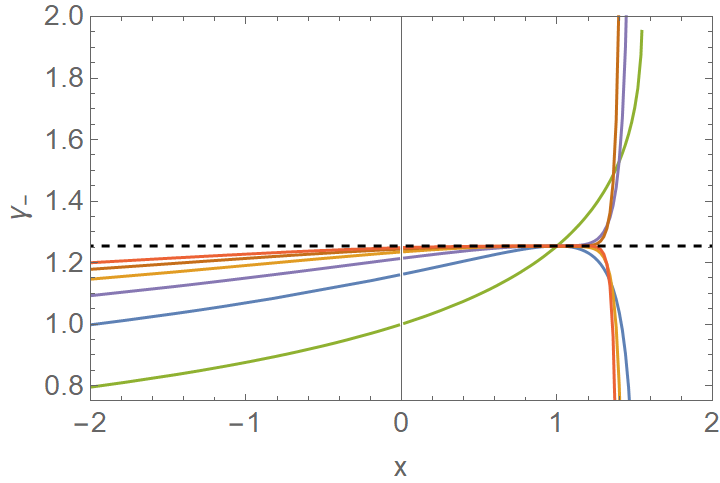} 
  \caption{
  Improved EFT expansion for the reduced binding momentum $\gamma_-$ of the shallow $T$-matrix pole up to N$^5$LO, as function of the resummation parameter $x$. Notation as in Fig. \ref{fig:table_of_phaseshifts}.  }
  \label{fig:poleconv}
\end{figure}

We can repeat the procedure with the improved LO. However, there are now two poles,
\begin{equation}
\gamma_\pm = \sum_{n=0}^{\infty}\gamma_\pm^{(n)} \, ,
\end{equation}
with
\begin{eqnarray}
    \gamma_{\pm}^{(0)} &=& \frac{1}{2\alpha x} 
    \left(1\pm\sqrt{1-4\alpha x}\right)\,,
    \\
    \gamma_{\pm}^{(1)} &=& \frac{(1-x) \alpha \gamma_{\pm}^{(0)2}}{1-2\alpha x\gamma_{\pm}^{(0)}}, 
    \\
    \gamma_{\pm}^{(2)} &=& 
    \frac{\gamma_{\pm}^{(1)2}}{\gamma_{\pm}^{(0)}} +\ldots,
    \\
    \gamma_{\pm}^{(3)} &=&  - \frac{\nu_2\gamma_{\pm}^{(0)4}}{1-2\alpha x\gamma_{\pm}^{(0)}} +\ldots,
    \\
    \gamma_{\pm}^{(4)} &=&  \frac{
    \gamma_{\pm}^{(3)}\gamma_{\pm}^{(1)}}{\gamma_{\pm}^{(0)}} +\ldots,
    \\
    \gamma_{\pm}^{(5)} &=&   \frac{\nu_3\gamma_{\pm}^{(0)6}}{1-2\alpha x\gamma_{\pm}^{(0)}} +\ldots,
\end{eqnarray}
where again we display only some of the terms at each order. The shallow pole $\gamma_{-}$ represents the physical bound state and converges to $\gamma$ with increasing order of the interaction, as seen in Fig. \ref{fig:poleconv}. Because the shape parameters contribute only at high orders, even at low orders $\gamma_-\approx \gamma$ when $x\approx 1$. The rate of convergence decreases as $x$ deviates significantly from 1. Still, each order improves on the previous one until $x\simeq 1.32$, where the corrections to the pole position become oscillating and non-convergent, with $|\gamma_{-}^{(1)}|\sim|\gamma_{-}^{(2)}|\sim\ldots\sim |\gamma_{-}^{(n)}|$. This effect resembles the \quotes{wall} found for few-body systems in Refs. \cite{Recchia:2022jih,Contessi:2023yoz}. Here this instability appears to happen for larger $x$ than in few-body systems, allowing for wider resummation.

The ``wall'' here is tied to the second pole. For $x$ large and negative, our improved LO is equivalent to the LO of an EFT with two low-energy poles~\cite{Habashi:2020ofb,Habashi:2020qgw,vanKolck:2022lqz}, with $\gamma_+$ representing a deeper virtual state. As $x$ increases, $\gamma_+$ gets deeper and emerges as a deep bound state. For $x=1/4 \alpha$, or $x\simeq 1.546$ for our value of $\alpha$, the two poles coincide at LO, $\gamma_{-}^{(0)}=\gamma_{+}^{(0)}$. For larger $x$, the two poles become complex, but on the upper half of the complex momentum plane and thus unphysical.

When one is primarily interested in the properties of the bound state, this expansion is cumbersome. It is much more practical to improve it by determining the one LO parameter not from the scattering length but from the binding momentum. Again, to be definite we compare the EFT with the toy underlying theory, taking for $\gamma$ the value given in Table \ref{tab:dimlessparam}. 

\begin{figure}[tb]
  \centering
  \includegraphics[width=0.85\linewidth]{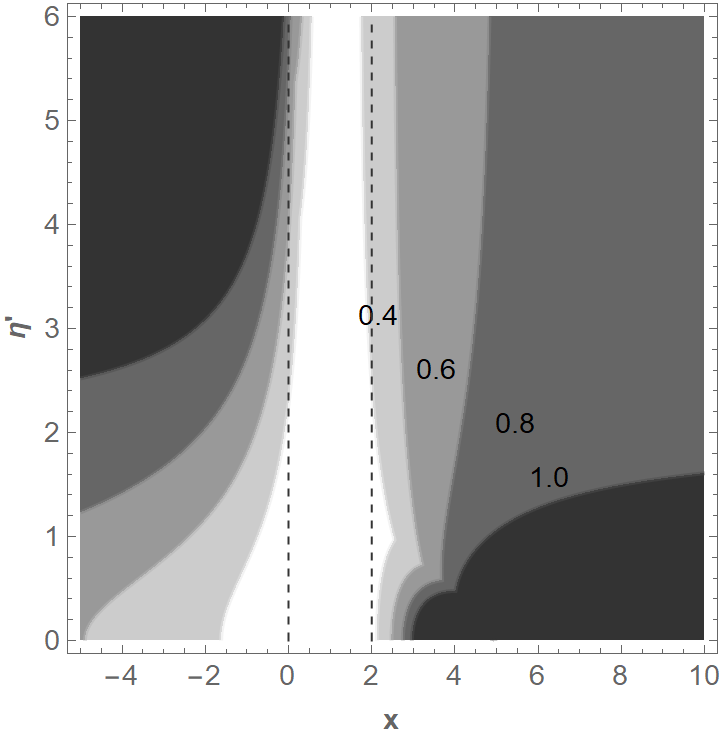} 
  \caption{Lowest-order convergence rate for the improved phase shifts in an expansion around the pole, $\xi_{\delta}'$ \eqref{eq:yetanotherxi}, as a function of the reduced momentum $\eta'$ and the resummation parameter $x$. The plot conventions are the same as Fig. \ref{fig:convergence}.
  }
  \label{fig:convergence_pole}
\end{figure}

\begin{figure}[tb]
  \centering
  \includegraphics[height=5cm]{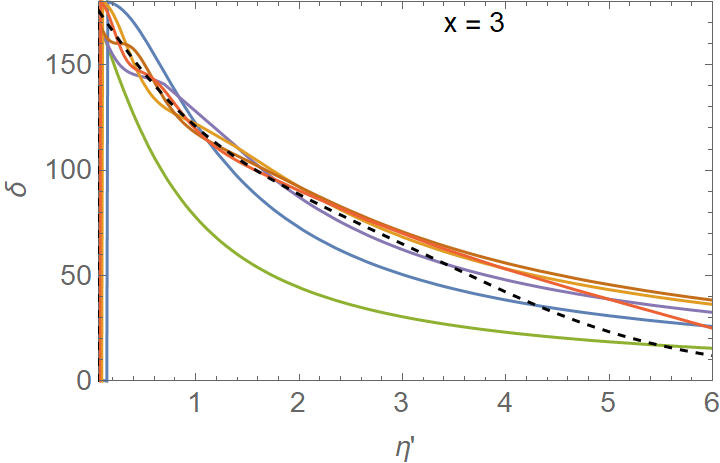} 
  \caption{Example of the convergence with order of the improved phase shifts (in degrees) as a function of the reduced momentum $\eta'$, for the dimensionless resummation parameter $x=3$. Notation as in Fig. \ref{fig:table_of_phaseshifts}.
  }
  \label{fig:phase_poleposition}
\end{figure}

The previous analysis can be repeated for the EFT expansion of the $T$ matrix around the pole position, which converges to Eq.~(\ref{eq:ERE-pole}). For this, we recalculate $\xi_{\delta}$ in the pole expansion, which is modified to
\begin{align}
    \xi_{\delta}'(x,\eta')=& \Bigg| 
    \frac{(1-x)\alpha' \eta' (1+\eta'^2)}{[1-x\alpha' (1+\eta'^2)]^2 +\eta'^2}   
    \nonumber\\ 
    &\times \left\{{\rm arccot}\left(\frac{1}{\eta'}\left[-1+\alpha' x \left(1+\eta'^2\right)\right]\right)\right\}^{-1}\Bigg|.
    \label{eq:yetanotherxi}
\end{align}
This function is displayed in Fig.~\ref{fig:convergence_pole}. Comparing with Fig. \ref{fig:convergence}, we notice that the convergence with respect to $x$ is worsened by this procedure, particularly at small values of $\eta'$ where there are now larger instabilities for $x \approx 2.5$. These instabilities manifest themselves in oscillations of the phase shifts which are clearly visible in Fig.~\ref{fig:phase_poleposition} already for $x\gtrsim 3$ (instead of $x\gtrsim 6$ as in the expansion around threshold). At zero momentum, $\xi_{\delta}'(\eta' \rightarrow 0)>1$ for $x \gtrsim 2.9648$. We conclude that, for scattering, the range of resummability of the theory is reduced in the case of a theory expanded around the pole position, both for positive and negative $x$. This might be due to the expansion point being shifted away from the scattering axis (real $\eta$) to $\eta'=i$.

Another difference between the two expansions is the behavior of the two poles introduced by the improved action.
In the expansion around the pole position, one of the poles is fixed by construction at each order while the second moves. For example, at LO, 
\begin{eqnarray}
\gamma_{-}^{(0)} &=& 1, \\
\gamma_{+}^{(0)} &=& \frac{1-x\,\alpha}{x\,\alpha}.
\end{eqnarray}
The two poles are on the imaginary axis for any (real) value of $x$, and $|\gamma_{+}| < |\gamma_{-}|$ for $x\gtrsim 2.465$. In contrast, when we expand around the scattering length, the two poles ultimately become unphysical with the same absolute value. The appearance of a second real, shallow pole on the positive imaginary axis does not appear to immediately affect the convergence of the $T$ matrix. However, its low-energy nature can be problematic, since, in principle, it describes an unphysical bound state within the convergence radius of the theory. Still, this is only a problem at the lowest orders of the power counting since the position of the crossing is not stable and shifts to larger $x$ when subleading orders are included.

\vspace{5mm}
\noindent
{\bf Charge form factor:}
We can improve our understanding of the convergence patterns when range corrections are resummed ($x \neq 0$) by considering the EFT expansion for other quantities besides the phase shifts. As an example, we look at the charge form factor of the two-body bound state.

The charge form factor describes how a bound state, here the deuteron, interacts with the electric field. It is related to the bound-state wave function in coordinate space, $\Psi(\vec{r})$, by
\begin{eqnarray}
  G_Q(\vec{q}) =
  \int d^3 r\,{|\Psi(\vec{r})|}^2\,e^{i \vec{q} \cdot \vec{r}/2}
  + \delta G_Q(\vec{q})
  \, ,
  \label{eq:FF}
\end{eqnarray}
with $\vec{q}$ the momentum of the photon. The first integral represents the contribution from the direct interaction of the photon with the point charge of the constituents, the so-called ``point'' form factor, while $\delta G_Q$ accounts for both one-body contributions arising from the constituent sizes and two-body current contributions. 

The point form factor depends solely on the bound-state wave function. As it is well known \cite{Mukhamedzhanov:2022zam}, at large distances
\begin{eqnarray}
 \Psi(\vec{r}) \simeq a_S \sqrt{\frac{\kappa}{2 \pi}}\,\frac{e^{-\kappa r}}{r}
 \label{eq:WFfinal}
\end{eqnarray}
has a dimensionless asymptotic normalization $a_S$ given by the residue of the $T$ matrix at the corresponding single pole \cite{Hu:1948zz},
\begin{eqnarray}
  T(\eta') = -i K'\frac{a_S^2}{\eta' -i} + f(\eta')\, ,
\end{eqnarray}
where $f(\eta')$ is regular at $\eta'=i$. In terms of the ERE,
\begin{eqnarray}
  a_S^2
  =  \left( 1 - 2 \alpha' + 4 \nu_2' - 6 \nu_3' +\ldots \right)^{-1}
\, .
\end{eqnarray}

In our toy underlying theory, the expansion of $a_S^{-2}$ is truncated at the $\nu_3'$ term. The wave function will differ from Eq. \eqref{eq:WFfinal} at distances comparable to the range $R$ of the potential, but at these distances two-body contributions to the form factor also exist. In order to focus on the EFT improvement of the wave function, we neglect one-body corrections. We will consider the case where the only effect of $\delta G_Q$ is to provide the two-body currents that cancel the short-range contributions of the wave function while ensuring charge conservation, $G_Q(0) = 1$. Introducing 
\begin{eqnarray}
  \theta' = \frac{q}{\kappa} \, ,
\end{eqnarray}
the evaluation of the form factor gives
\begin{eqnarray}
  G_Q(\theta') = 1 - {a_S^2}\left[1- \frac{4}{\theta'}\,{\arctan}\,\left( \frac{\theta'}{4} \right)\right] \, ,
\end{eqnarray}
where the nonanalytic term comes from the integral in Eq. \eqref{eq:FF} and the constants from $\delta G_Q$.

As we are not including currents beyond those necessary for charge conservation, the EFT expansion of the charge form factor~\cite{Chen:1999tn} only involves the expansion of the reduced asymptotic normalization $a_S^2$ in powers of $QR$. The improved expansion corresponds to
\begin{eqnarray}
  a_S^2 = \frac{1}{1 - 2x\alpha'} 
  \left[ {1 - \frac{2(1-x)\alpha'}{1 - 2 x \alpha'} + \dots} \right] \, .
\label{eq:aS2expansion}
\end{eqnarray}

\begin{figure*}
  \centering
  \begin{tabular}{cc}
    \multicolumn{2}{c}{ \includegraphics[width=0.6\textwidth]{Legend.png}} \\
    \includegraphics[width=0.45 \linewidth]{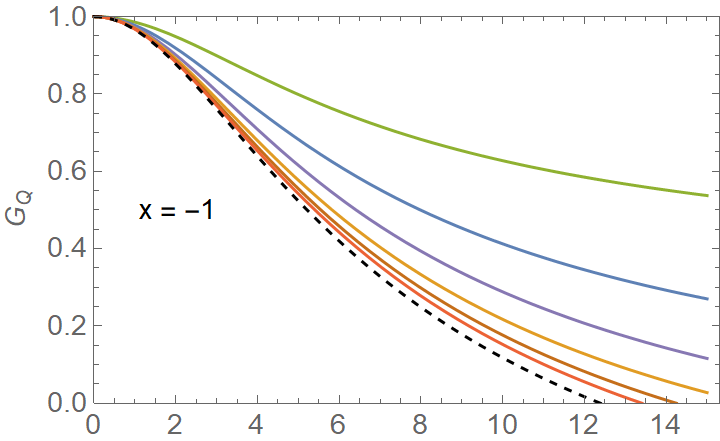}  & \includegraphics[width=0.45 \linewidth]{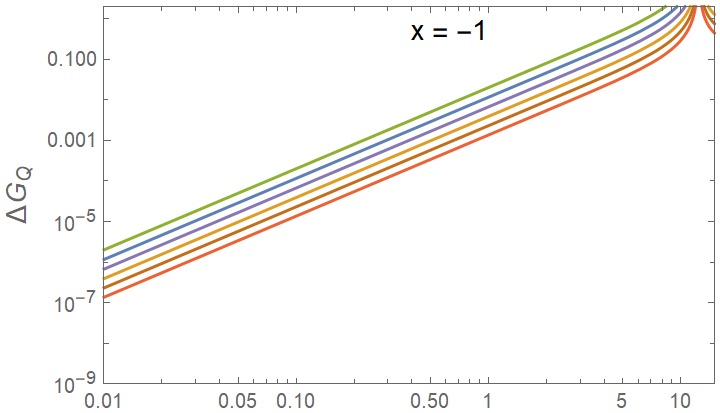}   \\
    \includegraphics[width=0.45 \linewidth]{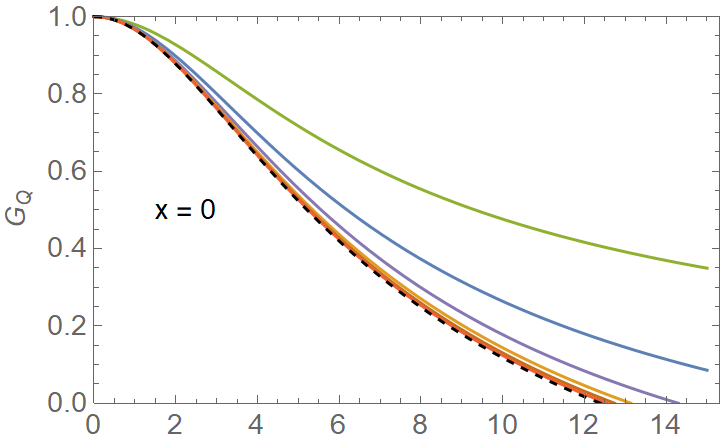}  & \includegraphics[width=0.45 \linewidth]{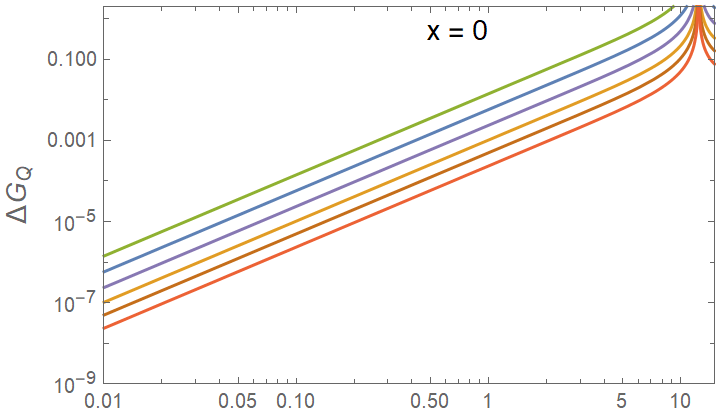}   \\
    \includegraphics[width=0.45 \linewidth]{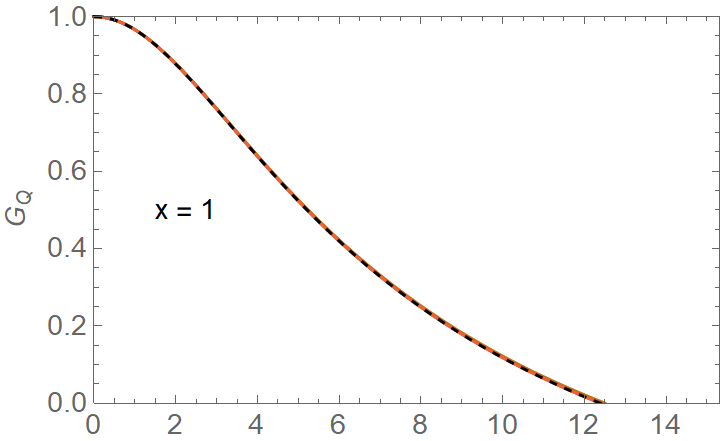}  & \includegraphics[width=0.45 \linewidth]{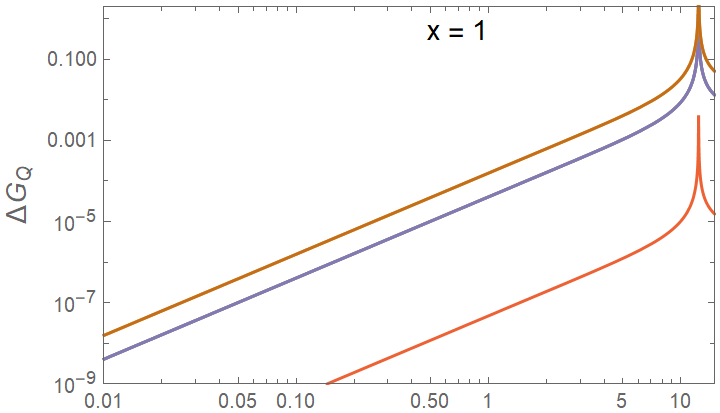}   \\
    \includegraphics[width=0.45 \linewidth]{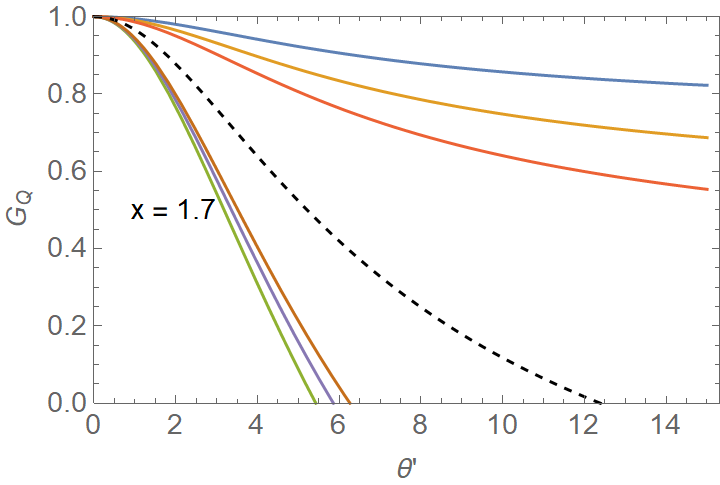}  & \includegraphics[width=0.45 \linewidth]{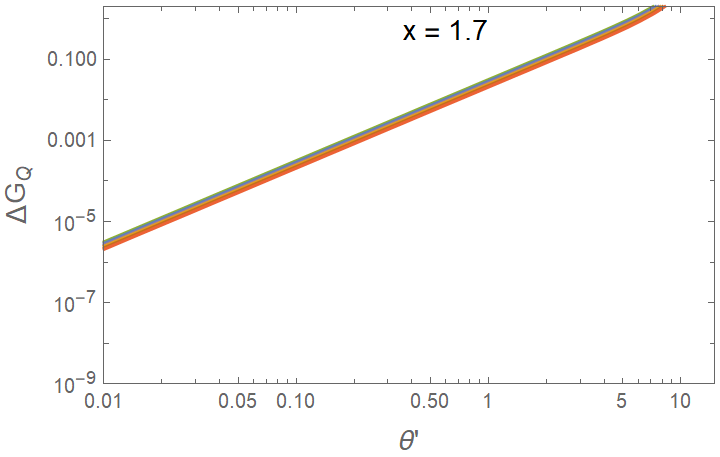}   
  \end{tabular}
  \caption{Improved EFT expansion of the bound-state (deuteron) electric form factor up to N$^5$LO for selected values of the resummation parameter $x$: $x=-1$ (top row), $x=0$ (second row), $x=1$ (third row), and $x=1.7$ (bottom row). On the left panels, the EFT and toy (dimensionless) form factors  are shown as functions of the photon momentum $\theta'$. On the right panels, the differences between EFT and toy form factors divided by the toy form factor ($\Delta G_Q$) are shown as functions of $\theta'$, both on logarithmic scales. Notation as in Fig. \ref{fig:table_of_phaseshifts}.
  }
        \label{fig:table_of_formfactors}
    \end{figure*}

The convergence of the improved expansion for the charge form factor is shown in Fig.~\ref{fig:table_of_formfactors} for various values of $x$. The left panels compare the six lowest orders with the underlying toy-theory result where $a_S$ is not expanded, while the right panels show the corresponding Lepage plots. Since the momentum dependence does not change with order, it has little impact on the convergence of the theory, which is dominated by the value of $x$ or, equivalently, by the convergence of the wave function. For $x\lesssim 1.732$, each order brings the EFT result closer to the underlying toy-theory result, with a decrease in the relative error shown in the Lepage plots. As expected, the fastest improvement occurs for $x=1$, when the effective range is fully incorporated at LO~\cite{Phillips:1999hh}, leaving only the small shape parameters to be reproduced at higher orders. Improvement deteriorates as $x$ deviates from 1. At $x=1.7$ it is still possible to discern a slow convergence, which disappears at $x \simeq 1.732$.

Figure \ref{fig:table_of_formfactors} provides a nice visualization of the convergence pattern with order but, because the expansion in the photon momentum does not come into play, the Lepage plots do not give an indication about the breakdown scale of the theory. This is a consequence of our neglect of subleading current effects accounted for in the $\delta G_Q(\vec{q})$ of Eq.~(\ref{eq:FF}), which can be expanded in powers of $q^2$. The convergence here reflects simply a change in the point charge radius, which is determined by $a_S^2$. The dramatic change in behavior occurs where the second term in Eq. \eqref{eq:aS2expansion} is equal to the first term: $x= (1+1/2\alpha')/2$, or $x \simeq 1.732$ for the effective-range value we are using. The reduced value in $x$ compared to scattering is partially due to a factor of 2 that arises from $|\Psi(\vec{r})|^2$. Although this is a somewhat artificial example, it illustrates how action improvement is observable dependent.

\vspace{5mm}
\noindent 
{\bf Conclusions:} We have examined the convergence of Short-Range EFT in the two-body sector when part of the effective range is resummed at LO. We characterize the resummed expansion with the dimensionless parameter $x$, which is the fraction of the physical effective range that is resummed at LO. In general we notice that the lowest orders in the EFT expansions of the scattering amplitude and the charge form factor quickly approach the \quotes{exact}  results of a toy underlying theory for values of $x$ between 0 and 1. Larger values seem to be acceptable as well, though the improvement deteriorates for $x\lesssim 0$ and $x\gtrsim 2$. Examples are the phase shifts, which converge rather slowly for $x=6$ if we expand around threshold, or $x=3$ if we expand around the bound-state pole. Not surprisingly, lower-order improvement is found to be maximal when the full effective range is resummed~\cite{Phillips:1999hh}. When we consider finite-cutoff effects in the resummed EFT, we see that they do not change the previous conclusions.

This resummation is not prescribed by the power counting for two-body systems with a single low-energy pole, or equivalently large scattering length and natural effective range~\cite{vanKolck:1998bw,Chen:1999tn}. Indeed, while improvement is observed at lowest orders, the resummation does not alter the fundamental convergence properties of the EFT expansion. Our simple toy underlying theory allow comparisons up to high orders, which in turn reveal that both resummed and non-resummed expansions converge to the same results within the same convergence radius, for a sensible choice of how much of the effective range is resummed at LO. The only difference is that for certain resummations the convergence at the lower orders of the expansion is faster, while for others it is slower. 

The resummation of part of the effective range brings in a second pole to the improved LO, which must remain far from the physical pole. This is very different from another class of systems when the effective range is also large in magnitude and there are two low-energy poles. In this case there is no choice but to include at LO also a short-range interaction that accounts for the effective range. The difficulty with including range effects exactly is the singularity of a two-derivative contact interaction. Renormalization constrains the effective range to be non-positive \cite{Phillips:1996ae,Beane:1997pk} and the pole positions in such a way that only one is a bound state on the upper half of the complex momentum plane \cite{Habashi:2020ofb,Habashi:2020qgw,vanKolck:2022lqz}. Resumming a positive effective range can be accomplished simply with a dimer ghost field, but that leads to problems in the three-body system \cite{Gabbiani:2001yh} which are absent for a negative effective range \cite{Griesshammer:2023scn}. Alternatively, one can employ at LO either a nonlocal interaction \cite{Beane:2021dab,Timoteo:2023dan} or a finite-range potential \cite{Contessi:2023yoz}. In either case, if only one pole is physical the challenge is to preserve the order-by-order convergence of the theory. 

The critical aspect of our improvement is that it complies with the perturbative nature of higher orders. In particular, the physical effective range remains an NLO effect, that is, it is not necessarily fitted at LO as reflected in the arbitrariness of $x$. The lower-order improvement, while not affecting the ultimate convergence of EFT, might provide a better starting point for many-body calculations. We have recently shown that this is the case for systems of bosons up to NLO \cite{Contessi:2023yoz}, where binding energies stop improving rather suddenly beyond $x=1$. Our two-body results here demonstrate the advantages and limitations of resummations in a simpler context. The improvement is seen to be observable dependent and to require that convergence be studied for each observable of interest. We intend to deploy a similar improvement to fermionic few-body systems next.

\vspace{5mm}
\noindent 
{\bf Acknowledgements:}
This work was supported in part
by the U.S. Department of Energy, Office of Science, Office of Nuclear Physics, under award DE-FG02-04ER41338 (UvK), the National Natural Science Foundation of China under Grants No. 12047503 and No. 12125507,  
the Fundamental Research Funds for the Central Universities and
the Thousand Talents Plan for Young Professionals (MPV).

\bibliographystyle{apsrev}
\bibliography{resummation.bib}

\end{document}